\documentclass[onecolumn,amssymb,amsmath,showpacs]{revtex4}

\usepackage{graphicx}
\usepackage{dcolumn}
\usepackage{bm}
\usepackage{lipsum}
\usepackage{amsmath,amsfonts,amssymb,mathrsfs,subfigure}
\usepackage{color,psfrag}

\def\bm{\textbf{m}}
\def\bh{\textbf{h}}
\def\be{\textbf{e}}
\def\bx{\textbf{x}}
\def\bw{\textbf{w}}
\def\bc{\textbf{c}}
\newcommand{\abs}[1]{\lvert#1\rvert}

\begin{document}

\preprint{APS/123-QED}

\title{Spin waves in ferromagnetic thin films}

\author{Zhiwei Sun}
 \affiliation{School of Mathematical Sciences, Soochow University, Suzhou, China}


\author{Jingrun Chen}
 \email{jingrunchen@suda.edu.cn}
  \affiliation{School of Mathematical Sciences, Soochow University, Suzhou, China}%
 \affiliation{Mathematical Center for Interdisciplinary Research, Soochow University, Suzhou, China}%

\date{\today}%

\begin{abstract}
A spin wave is the disturbance of intrinsic spin order in magnetic materials. In this paper, a spin wave in the 
Landau-Lifshitz-Gilbert equation is obtained based on the assumption that the spin wave maintains its shape while it propagates at a constant velocity.
Our main findings include:
(1) in the absence of Gilbert damping, the spin wave propagates at a constant velocity with the increment proportional to the strength of the magnetic field;
(2) in the absence of magnetic field, at a given time the spin wave converges exponentially fast to its initial profile as the damping parameter goes to zero and in the long time the relaxation dynamics of the spin wave converges exponentially fast to the easy-axis direction with the exponent proportional to the damping parameter; (3) in the presence of both Gilbert damping and magnetic field, the spin wave converges to the easy-axis direction exponentially fast
at a small timescale while propagates at a constant velocity beyond that. These provides a comprehensive understanding of spin waves
in ferromagnetic materials.
\end{abstract}

\pacs{05.45.Yv, 75.70.-i, 75.78.-n}%

\maketitle

\section{Introduction}

A spin wave is the disturbance of intrinsic spin order in magnetic materials. It is usually excited using magnetic fields
and offers unique properties such as charge-less propagation and high group velocities, which are important for signal 
transformations and magnetic logic applications~\cite{Kostylev:2005, Kajiwara:2010, Chumak:2015, Woo:2017, Chumak:2017, Langer:2019}.
 
The propagation of spin waves is described by the Landau-Lifshitz-Gilbert (LLG) equation \cite{LandauLifshitz:1935, Gilbert:1955}
in the dimensionless form
\begin{equation}\label{eqn:llg}
\bm_t = - \bm \times \bh - \alpha \bm \times (\bm \times \bh),
\end{equation}
where the magnetization $\bm = (m_1, m_2, m_3)^T$ is a three dimensional vector with unit length, $\alpha$ is the Gilbert
damping parameter. The effective field $\bh$ includes the exchange term, the anisotropy term with easy axis along the x-axis and
the anisotropy constant $q$, and the external field
\begin{equation}\label{eqn:h}
\bh = \Delta \bm + qm_1\be_1 + h_{\textrm{ext}}\be_1.
\end{equation}
$h_{\textrm{ext}}$ is the strength of the external field applied along the x-axis with $\be_1$ the unit vector.
This model is often used to describe the magnetization dynamics in ferromagnetic thin films.

From a theoretical perspective, a spin wave is known as a solitotary wave, which appears as the solution
of a weakly nonlinear dispersive partial differential equation. 
In LLG equation \eqref{eqn:llg}-\eqref{eqn:h}, a soliton is caused by the cancellation of nonlinear and dispersive effects
in the magnetic material. Solitons are of interests for quite a long time \cite{Nakamura:1974, Mikeska:1977, Lakshmanan:1984, Magyari:1986, Guo:2007}.
Most of works consider the one dimensional case and drop the damping term \cite{Nakamura:1974, Mikeska:1977, Guo:2007}.
In \cite{Lakshmanan:1984}, using the stereographic projection, the authors found that the presence of Gilbert damping was merely
a rescaling of time by a complex constant. However, this was found to be valid only for a single spin in a constant magnetic field \cite{Magyari:1986}.

In this work, we give a comprehensive study of an explicit spin wave in the LLG equation. Our starting point is that the spin wave maintains its shape while it propagates at a constant velocity and the derivation is based on the generalization of the method of characteristics. The main findings are:
(1) in the absence of Gilbert damping, the spin wave propagates at a constant velocity with the increment proportional to the strength of the magnetic field;
(2) in the absence of magnetic field, at a given time the spin wave converges exponentially fast to its initial profile as the damping parameter goes to zero and in the long time the relaxation dynamics of the spin wave converges exponentially fast to the easy-axis direction with the exponent proportional to the damping parameter; (3) in the presence of both Gilbert damping and magnetic field, the spin wave converges to the easy-axis direction exponentially fast
at a small timescale while propagates at a constant velocity beyond that.

\section{Derivation and results}

As mentioned above, we start with the assumption that a spin wave maintains its shape while it propagates at a constant velocity.
This can be seen from the method characteristics in simple situations.

In 1D when $\alpha = q = h_{\textrm{ext}} = 0$, one can check that
\begin{equation}\label{eqn:soliton1d}
\bm(x,t) =
\left(
\begin{array}{c}
\cos \theta_0 \\
\sin \theta_0 \cos \left( \frac{c}{\cos \theta_0}(x+ct) \right)  \\
\sin \theta_0 \sin \left( \frac{c}{\cos \theta_0}(x+ct) \right)  \\
\end{array}
\right)
\end{equation}
solves $\bm_t = - \bm \times \bm_{xx}$. Here $\theta_0$ is determined by
the initial condition and $u=x+ct$ is the characteristic line. \eqref{eqn:soliton1d} provides
a solitary solution with the traveling speed $c$. A detailed derivation of \eqref{eqn:soliton1d}
can be found in Chapter 2 of \cite{Guo:2007}.

A generalization of the method of characteristics yields a spin wave to $\bm_t = - \bm \times \Delta\bm$
\begin{equation}\label{eqn:soliton3d}
\bm(\bx,t) =
\left(
\begin{array}{c}
\cos \theta_0 \\
\sin \theta_0 \cos  \frac{v}{\cos \theta_0}  \\
\sin \theta_0 \sin  \frac{v}{\cos \theta_0}  \\
\end{array}
\right),
\end{equation}
where $v = c_1 x + c_2 y + c_3 z + (c_1^2+c_2^2+c_3^2)t = \bc\cdot\bx + (\bc\cdot\bc)t$ with $\bc =(c_1, c_2, c_3)^T$. 
The speed field is $\bc$ with magnitude $\abs{\bc}$.
Actually, both \eqref{eqn:soliton1d} and \eqref{eqn:soliton3d} can be rewritten as
\begin{equation}\label{eqn:assump1}
\bm(\bx,t) =
\left(
\begin{array}{c}
\cos \theta_0 \\
\sin \theta_0 \cos  \left( \bw_0\cdot\bx + \varphi(t) \right)  \\
\sin \theta_0 \sin  \left( \bw_0\cdot\bx + \varphi(t) \right)  \\
\end{array}
\right),
\end{equation}
where $\bw_0 = \bc/\cos\theta_0$ and $\varphi(t) = \left(\abs{\bc}^2/\cos\theta_0\right) t$.

\eqref{eqn:soliton1d}-\eqref{eqn:assump1} are obtained in the absence of Gilbert damping.
In order to take the Gilbert damping and the other terms in \eqref{eqn:h} into account, we make an ansatz
for the spin wave profile in the following form
\begin{equation}\label{eqn:assump2}
\bm(\bx, t) =
\left(
\begin{array}{c}
\cos \theta(t) \\
\sin \theta(t) \cos  \left(\bw_0\cdot \bx + \varphi(t)\right)  \\
\sin \theta(t) \sin  \left(\bw_0\cdot \bx + \varphi(t)\right)  \\
\end{array}
\right),
\end{equation}
where $\theta$ and $\varphi$ are independent of $\bx$ and only depend on $t$.

Substituting \eqref{eqn:assump2} into \eqref{eqn:h} and \eqref{eqn:llg} and denoting $\bw_0\cdot \bx + \varphi(t)$ by $u(\bx,t)$, we have
\begin{equation*}
\begin{split}
\bh =
\left(
\begin{array}{c}
0 \\
- \abs{\bw_0}^2 \sin \theta \cos u(\bx,t) \\
- \abs{\bw_0}^2 \sin \theta \sin u(\bx,t) \\
\end{array}
\right)
+ q
\left(
\begin{array}{c}
\cos \theta \\
0 \\
0 \\
\end{array}
\right)
+ h_{\textrm{ext}}
\left(
\begin{array}{c}
1 \\
0 \\
0 \\
\end{array}
\right),
\end{split}
\end{equation*}
\begin{equation*}
\begin{split}
\bm \times \bh =
\left(
\begin{array}{c}
0 \\
\abs{\bw_0}^2 \sin \theta \cos \theta \sin u(\bx,t) \\
- \abs{\bw_0}^2 \sin \theta \cos \theta \cos u(\bx,t) \\
\end{array}
\right)
+ q
\left(
\begin{array}{c}
0 \\
\sin \theta \cos \theta \sin u(\bx,t) \\
- \sin \theta \cos \theta \cos u(\bx,t) \\
\end{array}
\right) 
+ h_{\textrm{ext}}
\left(
\begin{array}{c}
0 \\
\sin \theta \sin u(\bx,t) \\
- \sin \theta \cos u(\bx,t) \\
\end{array}
\right) ,
\end{split}
\end{equation*}
\begin{equation*}
\begin{split}
\bm \times ( \bm \times \bh ) =&
\left(
\begin{array}{c}
- \abs{\bw_0}^2 \sin^2\theta \cos\theta \\
\abs{\bw_0}^2 \sin\theta \cos^2\theta \cos u(\bx,t) \\
\abs{\bw_0}^2 \sin\theta \cos^2\theta \sin u(\bx,t) \\
\end{array}
\right)
+ q
\left(
\begin{array}{c}
- \sin^2\theta \cos\theta \\
\sin\theta \cos^2\theta \cos u(\bx,t) \\
\sin\theta \cos^2\theta \sin u(\bx,t) \\
\end{array}
\right)
+ h_{\textrm{ext}}
\left(
\begin{array}{c}
- \sin^2\theta \\
\sin\theta \cos\theta \cos u(\bx,t) \\
\sin\theta \cos\theta \sin u(\bx,t) \\
\end{array}
\right),
\end{split}
\end{equation*}
and
\begin{equation*}
\bm_t =
\left(
\begin{array}{c}
- \theta_t \sin \theta \\
\theta_t \cos \theta \cos u(\bx,t)
- \varphi_t \sin \theta \sin u(\bx,t) \\
\theta_t \cos \theta \sin u(\bx,t)
+ \varphi_t \sin \theta \cos u(\bx,t) \\
\end{array}
\right).
\end{equation*}

After algebraic simplifications, we arrive at
\begin{equation}\label{eqn:ode}
\left\{
\begin{array}{ll}
\theta_t = - \alpha ( \abs{\bw_0}^2 + q )\sin\theta \cos\theta - \alpha h_{\textrm{ext}}\sin\theta \\
\varphi_t = ( \abs{\bw_0}^2 + q ) \cos \theta + h_{\textrm{ext}}
\end{array}
\right..
\end{equation}

\subsection{The absence of Gilbert damping}

When $\alpha = 0$, we have $\theta = \theta_0$ and $\varphi = \left( ( \abs{\bw_0}^2 + q ) \cos \theta_0 + h_{\textrm{ext}}\right)t$.
Therefore we have the solution
\begin{equation}\label{eqn:solution1}
\bm =
\left(
\begin{array}{c}
\cos \theta_0 \\
\sin \theta_0 \cos  \left(\bw_0\cdot \bx + t\left( \abs{\bw_0}^2\cos \theta_0 + q  \cos \theta_0 + h_{\textrm{ext}}\right)\right)  \\
\sin \theta_0 \sin  \left(\bw_0\cdot \bx + t\left( \abs{\bw_0}^2\cos \theta_0 + q  \cos \theta_0 + h_{\textrm{ext}}\right)\right)  \\
\end{array}
\right).
\end{equation}
Note that this recovers \eqref{eqn:assump1} when $q=0$ and $h_{\textrm{ext}}=0$. 
It is easy to see that the spin wave \eqref{eqn:solution1} propagates at a constant velocity.
The increment of the velocity field is
$q\cos\theta_0\frac{\bw_0}{\abs{\bw_0}^2}$ with magnitude $\frac{\abs{q\cos\theta_0}}{\abs{\bw_0}}$, due to the magnetic anisotropy.
The increment of the velocity field is
$h_{\textrm{ext}}\frac{\bw_0}{\abs{\bw_0}^2}$ with magnitude $\frac{\abs{h_{\textrm{ext}}}}{\abs{\bw_0}}$, due to the magnetic field.

\subsection{The absence of magnetic field}

When $ h_{\textrm{ext}} = 0$, \eqref{eqn:ode} reduces to
\begin{equation}\label{eqn:ode0}
\left\{
\begin{array}{ll}
\theta_t = - \alpha ( \abs{\bw_0}^2 + q )\sin\theta \cos\theta \\
\varphi_t = ( \abs{\bw_0}^2 + q ) \cos \theta
\end{array}
\right..
\end{equation}

For the first equation in \eqref{eqn:ode0}, assuming $0\leq \theta_0 < \pi/2$, by separation of variables, we have
\begin{equation*}
\alpha( \abs{\bw_0}^2 + q ) t + C_1 = \ln \cot\theta,
\end{equation*}
where $C_1$ is a constant determined by the initial condition. 

Denote $\tilde{t} = \alpha( \abs{\bw_0}^2 + q )t + C_1$.
It follows that
\begin{equation}\label{solution:theta}
\tan\theta = e^{-\tilde{t}},
\end{equation}
from which one has
\begin{gather}
\label{eqn:costheta}
\cos\theta = \frac{1 }{\sqrt{1 + e^{-2\tilde{t}}}}, \\
\label{eqn:sintheta}
\sin\theta = \frac{1 }{\sqrt{1 + e^{2\tilde{t}}}}.
\end{gather}
When $t=0$, \eqref{eqn:costheta} turns to
\begin{equation}\label{C1}
\cos\theta_0 =  \frac{1 }{\sqrt{1 + e^{-2C_1}}},
\end{equation}
from which we can determine $C_1$ by the initial condition $\theta_0$.

As for $\varphi$, from the second equation in \eqref{eqn:ode0}, one has that
\begin{equation*}
\frac{\mathrm{d}\varphi}{\mathrm{d}\theta} = \frac{\mathrm{d}\varphi}{\mathrm{d}t} \cdot \frac{\mathrm{d}t}{\mathrm{d}\theta} = - \frac{1}{ \alpha \sin\theta}.
\end{equation*}
Therefore
\begin{equation}\label{eqn:varphi}
\alpha \varphi = -\int \frac{\mathrm{d}\theta}{\sin\theta}
= \frac{1}{2}\ln \left( \frac{1+\cos\theta}{1-\cos\theta} \right) + C_2
= \ln \cot\frac{1}{2}\theta + C_2,
\end{equation}
where
\[
	C_2  = - \frac{1}{2}\ln \left( \frac{1+\cos\theta_0}{1-\cos\theta_0} \right).
\]

Substituting \eqref{eqn:costheta} and \eqref{C1} into \eqref{eqn:varphi} yields
\begin{equation}\label{solution:varphi}
\varphi = \frac{1}{\alpha}\ln\left(\frac{e^{\tilde{t}}+\sqrt{e^{2\tilde{t}}+1}} {e^{C_1}+\sqrt{e^{2C_1}+1}}\right)
=\frac{1}{\alpha}\left(\ln\cot\frac{1}{2}\theta -\ln\cot\frac{1}{2}\theta_0\right).
\end{equation}

In short summary, the spin wave when $\alpha\neq 0$ takes the form 
\begin{equation}\label{eqn:solution2}
\bm = \frac{1 }{\sqrt{1 + e^{2\tilde{t}}}}
\left(
\begin{array}{c}
e^{\tilde{t}} \\
\cos \left( \bw_0\cdot \bx + \varphi \right) \\
\sin \left( \bw_0\cdot \bx + \varphi \right) \\
\end{array}
\right).
\end{equation}

The above derivation is valid when $0\leq \theta_0 < \pi/2$. If $\pi/2 < \theta_0 \leq \pi$, we choose
the other solution of \eqref{eqn:costheta}
\begin{equation}\label{eqn:costheta1}
\cos\theta = -\frac{1}{\sqrt{1 + e^{-2\tilde{t}}}},
\end{equation}
and
\begin{equation*}
\varphi = - \frac{1}{\alpha}\ln\left(\frac{e^{\tilde{t}}+\sqrt{e^{2\tilde{t}}+1}} {e^{C_1}+\sqrt{e^{2C_1}+1}}\right).
\end{equation*}
\eqref{eqn:solution2} remains unchanged.

When $\alpha \rightarrow 0$, we have $\tilde{t} \rightarrow C_1$ and
\begin{equation*}
\lim\limits_{\alpha \rightarrow 0} \bm = \frac{1 }{\sqrt{1 + e^{2C_1}}}
\left(
\begin{array}{c}
e^{C_1} \\
\cos \left( \bw_0\cdot \bx + \lim\limits_{\alpha \rightarrow 0}\varphi \right) \\
\sin \left( \bw_0\cdot \bx + \lim\limits_{\alpha \rightarrow 0}\varphi \right) \\
\end{array}
\right).
\end{equation*}
By L'Hospital's rule, one has that
\begin{equation}\label{eqn:varphi0}
\lim\limits_{\alpha \rightarrow 0}\varphi 
= \lim\limits_{\alpha \rightarrow 0} \frac{\mathrm{d}}{\mathrm{d}\alpha} 
\left( \ln\left(\frac{e^{\tilde{t}}+\sqrt{e^{2\tilde{t}}+1}} {e^{C_1}+\sqrt{e^{2C_1}+1}}\right) \right)
=\lim\limits_{\alpha \rightarrow 0} \frac{e^{\tilde{t}}}{\sqrt{e^{2\tilde{t}}+1}} (\abs{\bw_0}^2+q)t
= \frac{e^{C_1}}{\sqrt{e^{C_1}+1}} (\abs{\bw_0}^2+q)t.
\end{equation}
Therefore it follows that
\begin{equation}\label{eqn:solution3}
\lim\limits_{\alpha \rightarrow 0} \bm =
\left(
\begin{array}{c}
\cos\theta_0 \\
\sin\theta_0 \cos \left( \bw\cdot \bx + \cos\theta_0 \left(\abs{\bw_0}^2+q\right)t \right) \\
\sin\theta_0 \sin \left( \bw\cdot \bx + \cos\theta_0 \left(\abs{\bw_0}^2+q\right)t \right) \\
\end{array}
\right).
\end{equation}
This is exactly the solution \eqref{eqn:solution1} when $h_{\textrm{ext}}=0$.

In addition, when $\alpha\rightarrow 0$, both $\theta$ and $\varphi$ converges exponentially fast to initial conditions; 
see equations \eqref{eqn:costheta}, \eqref{eqn:sintheta}, \eqref{C1}, and \eqref{eqn:varphi0}.
Therefore, at a given time $t$, \eqref{eqn:solution3} converges exponentially fast to the initial spin wave \eqref{eqn:solution1} when $h_{\textrm{ext}}=0$
with the exponent proportional to the damping parameter $\alpha$. 
Moreover, in the long time, i.e., when $t\rightarrow +\infty$, $\tilde{t}\rightarrow + \infty$ and $\theta\rightarrow 0$, 
\eqref{eqn:solution2} converges to $(1,0,0)^T$ (the easy-axis direction) exponentially fast with the rate proportional to the damping parameter $\alpha$.
When $\pi/2< \theta_0\leq \pi$, from \eqref{eqn:costheta1}, we have that \eqref{eqn:solution2} converges to $(-1,0,0)^T$ (again the easy-axis direction) exponentially fast with the rate proportional to the damping parameter $\alpha$.

It is easy to check that the right-hand side of \eqref{eqn:solution3} is
the solution of \eqref{eqn:llg} when $h_{\textrm{ext}}=0$ with the initial condition $\theta_0=\pi/2$.
Therefore, Gilbert damping does not have any influence on magnetization dynamics in this case.

In \cite{Lakshmanan:1984}, the authors used the stereographic 
projection and observed that the effect of Gilbert damping was only a rescaling of time by a complex constant.
However, this was latter found to be valid only for a single spin in a constant magnetic field \cite{Magyari:1986}.
Our result provides an explicit characterization of magnetization dynamics in the presence of Gilbert damping.

\subsection{The presence of both Gilbert damping and magnetic field}

It is difficult to get the explicit solution of \eqref{eqn:ode} in general. To understand the magnetization dynamics,
we use the method of asymptotic expansion. For small external magnetic field, $\theta$ and $\varphi$ admit the following expansions
\begin{align*}
	\theta(t,h_{\textrm{ext}}) & = \theta^0(t) + \theta^1(t)h_{\textrm{ext}} + \theta^2(t)h_{\textrm{ext}}^2 + \cdots,\\
	\varphi(t,h_{\textrm{ext}}) & = \varphi^0(t) + \varphi^1(t)h_{\textrm{ext}} + \varphi^2(t)h_{\textrm{ext}}^2 + \cdots.
\end{align*}
Therefore one has that 
\begin{align}\label{theta expantion}
	\theta_t(t,h_{\textrm{ext}}) & = \theta^0_t(t) + \theta^1_t(t)h_{\textrm{ext}} + \theta^2_t(t)h_{\textrm{ext}}^2 + \cdots,\\
	\label{varphi expantion}
	\varphi_t(t,h_{\textrm{ext}}) & = \varphi^0_t(t) + \varphi^1_t(t)h_{\textrm{ext}} + \varphi^2_t(t)h_{\textrm{ext}}^2 + \cdots.
\end{align}

On the other hand, from \eqref{eqn:ode}, it follows that
\begin{align}\label{eqn:theta expantion}
	\theta_t & = - \alpha ( \abs{\bw_0}^2 + q )\sin\theta^0 \cos\theta^0 
	 - \left(\alpha ( \abs{\bw_0}^2 + q )\theta^1\cos 2\theta^0 + \alpha\sin\theta^0 \right)h_{\textrm{ext}} + \cdots,\\
	 \label{eqn:varphi expantion}
\varphi_t & = ( \abs{\bw_0}^2 + q ) \cos \theta^0  + \left(- ( \abs{\bw_0}^2 + q )\theta^1\sin\theta^0 + 1\right) h_{\textrm{ext}} + \cdots.
\end{align}
Combining \eqref{theta expantion} and \eqref{varphi expantion} with \eqref{eqn:theta expantion} and \eqref{eqn:varphi expantion}, for the zero-order term, one has
\begin{equation}\label{eqn:zero order}
	\left\{
	\begin{array}{ll}
	\theta_t^0 = - \alpha ( \abs{\bw_0}^2 + q )\sin\theta^0 \cos\theta^0 \\
	\varphi_t^0 = ( \abs{\bw_0}^2 + q ) \cos \theta^0
	\end{array}
	\right.,
\end{equation}
which recovers \eqref{eqn:ode0} with solution \eqref{solution:theta} and \eqref{solution:varphi}.

As for the first-order term, one has that
\begin{equation}\label{eqn:first order}
\left\{
\begin{array}{ll}
	\theta_t^1 = - \alpha ( \abs{\bw_0}^2 + q )\theta^1\cos 2\theta^0 - \alpha\sin\theta^0 \\
	\varphi_t^1 = - ( \abs{\bw_0}^2 + q )\theta^1\sin\theta^0 + 1
\end{array}
\right..
\end{equation}
Using variation of parameters, one can assume $\theta^1 = \frac{C(t)}{e^{\tilde{t}} + e^{-\tilde{t}}}$ and it follows that
\begin{equation*}
	C'(t) = - \alpha (e^{\tilde{t}} + e^{-\tilde{t}})\sin\theta^0
	      = - \alpha (\tan\theta^0 + \tan^{-1}\theta^0)\sin\theta^0.
\end{equation*}
Since
\begin{equation*}
\begin{split}
	 \int - \alpha \tan\theta^0 \sin\theta^0 \mathrm{d}t
	=&\int - \alpha ( \abs{\bw_0}^2 + q )\sin\theta^0\cos\theta^0 *(\abs{\bw_0}^2 + q)^{-1}\frac{\sin\theta^0}{\cos^2\theta^0}\mathrm{d}t\\
	=& \int (\abs{\bw_0}^2 + q)^{-1}\frac{\sin\theta^0}{\cos^2\theta^0}\mathrm{d}\theta^0\\
	=&(\abs{\bw_0}^2 + q)^{-1}\frac{1}{\cos\theta^0},
\end{split}
\end{equation*}
and
\begin{equation*}
\begin{split}
	\int - \alpha \tan^{-1}\theta^0 \sin\theta^0 \mathrm{d}t
	=& \int - \alpha \cos\theta^0 \mathrm{d}t\\
	=& - \alpha(\abs{\bw_0}^2 + q)^{-1}\varphi^0,
\end{split}
\end{equation*}
one can get $C(t)=(\abs{\bw_0}^2 + q)^{-1} (\frac{1}{\cos\theta^0} - \alpha\varphi^0)$, and it follows that
\begin{equation}\label{theta^1}
\begin{split}
	\theta^1 =  (\abs{\bw_0}^2 + q)^{-1} (\sin\theta^0- \alpha\sin\theta^0\cos\theta^0\varphi^0).
\end{split}
\end{equation}

Substituting the first equation in \eqref{eqn:zero order} into the second equation in \eqref{eqn:first order}, one has
\begin{equation*}
\begin{split}
	 &-\int ( \abs{\bw_0}^2 + q )\theta^1\sin\theta^0 \mathrm{d}t\\
	 =&\frac{1}{\alpha}\int \frac{\theta^1}{\cos\theta^0} \mathrm{d}\theta^0\\
	 =&\frac{1}{\alpha(\abs{\bw_0}^2 + q)}\int \tan\theta^0 - \sin\theta^0(\ln\cot\frac{1}{2}\theta^0+C_2) \mathrm{d}\theta^0 \quad (\mathrm{using\; \eqref{theta^1}})\\
	 =&-t +\frac{1}{\abs{\bw_0}^2 + q}
	 (\varphi^0\cos\theta^0 +\alpha^{-1}C_1),
\end{split}
\end{equation*}
and thus
\begin{equation}
	\varphi^1 = \frac{1}{\abs{\bw_0}^2 + q}
	(\varphi^0\cos\theta^0 +\alpha^{-1}C_1).
\end{equation}

Therefore, when $h_{\textrm{ext}}$ is small, it has the approximate solution
\begin{equation}
\left\{
\begin{array}{ll}
\theta^* = \theta^0 +(\abs{\bw_0}^2 + q)^{-1} (\sin\theta^0- \alpha\sin\theta^0\cos\theta^0\varphi^0)h_{\textrm{ext}} \\
\varphi^* = \varphi^0 + (\abs{\bw_0}^2 + q)^{-1}
(\varphi^0\cos\theta^0 +\alpha^{-1}C_1)h_{\textrm{ext}}
\end{array}
\right.
\end{equation}
with $\theta^0$ and $\varphi^0$ satisfying \eqref{eqn:ode0}.

From \eqref{solution:theta} and \eqref{solution:varphi}, $\theta^0$ converges exponentially fast to the easy-axis direction, while $\varphi^0$ grows linearly. Therefore, from \eqref{theta^1}, $\theta^1$ converges
exponentially fast to $0$ as well with a larger exponent. This relaxation dynamics happens at a small timescale.

Meanwhile, from \eqref{eqn:first order}, the difference between $\varphi^*$ and $\varphi^0$ satisfies
\begin{equation}\label{eqn:varphidiff}
\varphi^*_t - \varphi_t^0 =  \varphi_t^1h_{\textrm{ext}} = - ( \abs{\bw_0}^2 + q )\theta^1\sin\theta^0h_{\textrm{ext}} + h_{\textrm{ext}}.
\end{equation}
Since $\theta^1\sin\theta^0$ converges to $0$ at a small timescale, the dynamics of $\varphi^* - \varphi^0$ is determined
by the external field at longer timescales. As a consequence, the increment of the velocity field is
$h_{\textrm{ext}}\frac{\bw_0}{\abs{\bw_0}^2}$ with magnitude $\frac{\abs{h_{\textrm{ext}}}}{\abs{\bw_0}}$.
This validates the Walker's ansatz \cite{Walker:1974} for a spin wave. 

When $\pi/2< \theta_0\leq \pi$ and the magnetic field is applied along the negative x-axis, and if $( \abs{\bw}^2 + q )| \cos\theta_0| \le  h_{\textrm{ext}}$, the result above will be correct. 

Note that $\theta_0=\pi/2$ does not fall into the above two cases since the magnetization dynamics will change the spin wave profile. In fact, as $t\rightarrow +\infty$, $\theta\rightarrow 0$ if the magnetic field is applied along the positive x-axis direction and $\theta\rightarrow \pi$ if the magnetic field is applied along the negative x-axis direction.

\section{Concluding remarks}

In this work, we study the magnetization dynamics in Landau-Lifshitz-Gilbert equation. 
By generalizing the method of characteristics, we are able to have an explicit characterization of spin wave dynamics in the presence of both
Gilbert damping and magnetic field. Gilbert damping drives the spin wave converge exponentially fast to the easy-axis direction with the exponent proportional
to the damping parameter at a small timescale and the magnetic field drives the spin wave propagate at a constant velocity at
longer timescales.

It will be of interests whether the technique developed here applies to the antiferromagnetic case \cite{Mikeska:1980, Baltz:2018}
and how rigorous the results obtained here can be proved from a mathematical perspective.

\section{Acknowledgements}

We thank Professor Yun Wang for helpful discussions. This work was partially supported by National Natural Science Foundation of China via grant 21602149 and 11971021. 

\bibliography{spintronics}

\end{document}